%Paper: hep-th/9211018
%From: benoitl@ERE.UMontreal.CA (Benoit Louis)
%Date: Wed, 4 Nov 92 16:22:47 -0500

%%%%%%%%%%%%%%%%%%%%%%%%%%%%%%%%%%%%%%%%%%%%%%%%%%%%%%%%%%%%%%%%%%
\documentstyle{amsppt}
\magnification=1200
\NoBlackBoxes
\TagsOnRight
\headline{{\hfill \folio}}
\nopagenumbers
\hsize=14.5truecm
\hoffset=2truecm
\vsize=23truecm
\settabs 16\columns

\font\titre=cmr17 at 17.3truept
\def\NS{Neveu-Schwarz }
\def\zhat{{\hat z}}
\def\zij#1#2{{\hat z_{#1#2}}}
\def\th{\theta}
\def\tij#1#2{{\theta_{#1#2}}}
\def\demi{{1\over2}}
\def\chphi#1{{\phi_h^{(#1)}}}
\def\jm{{J_-}}
\def\jo{{J_0}}
\def\jp{{J_+}}
\def\jpdeux{{J_+^2}}
\def\monstre{{(1+2t\jpdeux)}}

\
\vskip 1truecm
\centerline{\titre FUSION AND THE NEVEU-SCHWARZ SINGULAR VECTORS\footnote"*"
{This
work has been supported in part by NSERC Canada and the Fonds FCAR pour l'aide
et le soutien \`a la recherche (Qu\'ebec).}}

\vskip 2truecm
\baselineskip=14truept
\centerline{Louis BENOIT\footnote"$\dag$"{Address starting January 1993:
Physics Department, Jadwin Hall, Princeton University, Princeton, NJ, USA
08544.}}
\centerline{\sl D\'epartement de physique and Centre de recherches
math\'ematiques,}
\centerline{\sl Universit\'e de Montr\'eal, CP 6128A, Montr\'eal, Qu\'ebec,
Canada H3C 3J7}
\bigskip
\centerline{Yvan SAINT-AUBIN\footnote"$\ddag$"
{Address during the academic year
1992-93: School of Mathematics, Institute for Advanced Study, Princeton, NJ,
USA 08540.}}
\centerline{\sl Centre de recherches math\'ematiques and D\'epartement de
 math\'ematiques et de statistique,}
\centerline{\sl Universit\'e de Montr\'eal, CP 6128A, Montr\'eal, Qu\'ebec,
Canada H3C 3J7}
\bigskip\bigskip
\centerline{CRM--1833}
\centerline{October 1992}
\vskip 1.5 truecm

\noindent ABSTRACT.\ \ {\sl Bauer, Di Francesco, Itzykson and Zuber proposed
recently an algorithm to construct all singular vectors of the Virasoro
algebra.
It is based on the decoupling of (already known) singular fields in the
fusion  process. We show how to extend their algorithm to the Neveu-Schwarz
superalgebra.}
\bigskip
\noindent RESUME.\ \ {\sl Bauer, Di Francesco, Itzykson and Zuber proposaient
r\'ecemment un algorithme pour construire tous
les vecteurs singuliers de l'alg\`ebre
de Virasoro.  Cet algorithme repose sur le d\'ecouplage de champs singuliers
(d\'ej\`a connus)  lors du processus de fusion.  Nous montrons comment
\'etendre leur algorithme \`a l'alg\`ebre de Neveu-Schwarz.}

\vfill\eject

\baselineskip=24truept

\subhead I. INTRODUCTION\endsubhead
\medskip

Explicit expressions for various objects have great importance in theoretical
physics.  One might think, for example, to the discovery of the one-instanton
solution by Belavin, Polyakov, Schwartz and Tyupkin
and its impact on non-perturbative effects in QCD.  A
similar role was played by $N$-solitons solutions of the Korteweg-de Vries
equation for the field
of integrable models.  In representation theory such
examples are numerous: from Racah formula for $SU(2)$ 6-$j$
coefficients to Weyl formula for characters of finite-dimensional
irreducible representations of simple Lie algebras, these expressions are now
common tools in modern physics.  Besides their computational uselfulness,
these expressions brought with their discovery a wealth of information
about their properties and their relationships with other often remote
concepts of the theory.

Singular vectors of the Virasoro algebra and of its supersymmetric extensions,
the Neveu-Schwarz and the Ramond superalgebras, play a central
role in conformal
quantum field theories.  For this reason efforts were made to discover their
expressions.  According to early works by Kac \cite1 and Feigin and Fuchs
\cite2, we know
that, for the Virasoro algebra, singular vectors arise at level $pq$ in Verma
modules $V_{(c,h)}$ whose highest weight $h$ lies on one of the parametrized
curves $h_{p,q}(t)=
{1\over4}(1-p^2)t^{-1}+{1\over 2}(1-pq)+{1\over 4}(1-q^2)t$,
where $p$ and $q$ are
two positive integers. The continuous parameter $t$ is related to the central
charge $c=13 + 6(t+t^{-1})$.  The authors \cite3 were able to give an explicit
expression for the singular vectors whenever $p$ or $q$ equals to 1.  Though
this expression is particularly simple, it does not provide any clue on a
possible generalization for general $p$ and $q$ nor does it establish any
relationship between the algebraic concept and the more physical quantities of
conformal quantum field theory (CQFT).  Exploiting the operator
product expansion
of CQFT, Bauer, di Francesco, Itzykson and Zuber \cite4 were able
to provide such
a relationship.  Their algorithm reproduces the singular vectors
given in \cite3
though it does not lead to any new general expressions for $p$ and
 $q$ different
than 1. Hence this new result is not important because it answers
the problem of
explicit expressions for singular vectors but because it links in a deep way
the very existence of these vectors (an algebraic concept) with the
consistency
of the operator product expansion (a physical and analytic concept).

The \NS and Ramond singular vectors are as intriguing as Virasoro ones.  In an
earlier communication \cite5 we gave an explicit expression for the \NS
singular vectors with one index equal to one. Due to the importance of the
results by Bauer {\sl et al}, we feel that it is worthwhile to
extend them to the
\NS supersymmetric extension.  This goal is accomplished in the present paper
and, as will be seen, the superfield
formulation plays here a significant role.
Consequently the Ramond case remains an open problem. We recognize however
that the \NS and the Ramond cases might be intimately connected.

The overall organization of the paper is as follows: notations and basic
results are gathered in the next section. Section 3 develops the core of the
algorithm and gives a non-trivial example.  Section 4 is devoted to the
proof of one of the main identities (eq.~(3.10)). Concluding remarks follow.

\bigskip\bigskip
\subhead II. SOME PARTICULAR SINGULAR VECTORS\endsubhead
\nobreak
\medskip
\nobreak
The purpose of this section is to gather the basic ingredients to extend to the
\NS algebra the algorithm proposed by Bauer, Di Francesco, Itzykson and
Zuber (herafter BdFIZ).  Though nothing is new
here, some of the results are presented in an unusual way that is more in line
with the current goal.  The three important points
to be discussed are: ({\sl i})
the \NS superalgebra and its singular vectors, ({\sl ii}) the particular case
of the singular vectors $\vert \psi_{1,q}\rangle$  and ({\sl iii}) the
differential operators associated with the generators $L_{-n}$ and $G_{-r}$,
$n, r>0$.

The energy-momentum tensor $T(Z)$ is the generator of superconformal
transformations in a superconformal quantum field theory.  Throughout we are
using the superfield formalism in which $Z$ stands for the pair $(z,\theta)$,
$z$ being the usual ``chiral'' coordinate $z=x+iy$ ($x,y\in\Bbb R$) and
$\theta$ a Grassman variable that allows expansion of superfields in bosonic
and fermionic parts.  The superfield $T(Z)$ is fermionic
 and has conformal weight
$3\over2$: $T(Z)=T_F(z)+\theta T_B(z)$. Its fermionic and bosonic modes
$$\align
T_F&=\sum_{n\in\Bbb Z}{1\over2} z^{-n-2}G_{n+{1\over2}}\\
T_B&=\sum_{n\in\Bbb Z}z^{-n-2}L_n\\
\endalign$$
span, together with the central element, the \NS algebra:
$$\aligned
[L_n, L_m]&=(n-m)L_{n+m} + {c\over 12}(n^2-1)n\delta_{n,-m}\\
[L_m, G_r]&=({m\over 2}-r)G_{m+r}\\
\{G_r, G_s\}&=2L_{r+s}+{c\over 3}(r^2-{1\over 4})\delta_{r,-s},\\
\endaligned \tag2.1$$
where $n,m\in\Bbb Z$ and $r,s\in\Bbb Z+{1\over2}$.
The complex number $c$ is the
central charge of the theory.

Verma modules $V_{(c,h)}$ are constructed from a highest
 weight vector $\vert h
\rangle$ with the following properties
$$\aligned
L_n\vert h\rangle&
=G_r\vert h\rangle=0\qquad\text{for}\quad n\ge 1, r\ge {1\over
2}\\
L_0\vert h\rangle&=h\vert h \rangle\endaligned\tag2.2
$$
where $h$ is a complex number.
The vectors $$L_{-m_1}
\dots L_{-m_k}G_{-r_1}\dots G_{-r_l}\vert h\rangle\tag2.3
$$
with
$$m_1\ge m_2\ge \dots \ge m_k\ge1\qquad \text{and}\qquad
r_1>r_2>\dots>{1\over2}
$$
form a basis for the vector space.
The vector (2.3) has weight $h+\sum_{i=1}^k n_i+\sum_{j=1}^l r_j$.
 We also say
that it has level $\sum_{i=1}^k n_i+\sum_{j=1}^l r_j$. A singular
 vector $\vert
v\rangle\in V_{(c,h)}$ is a vector that satisfies the defining
properties of the
highest weight vector:
$$\align
L_n\vert v\rangle&=G_r\vert v\rangle=0\qquad\text{for}\quad n\ge 1,
 r\ge {1\over
2}\\
L_0\vert v\rangle&=(h+n)\vert v \rangle\endalign
$$
but has a weight strictly larger than $h$: $n\ge 1$.  Is is easily
seen that, by
acting with the $L_{-n}$ and $G_{-r}$, $n,r>0$, one can construct
from $\vert v\rangle$
an invariant submodule of $V_{(c,h)}$.  The physically natural inner product
on $V_{(c,h)}$ is defined by $\langle h\vert h\rangle=1$ and $L_n^\dag=L_{-n}$
and $G_r^\dag=G_{-r}$.  Hence a singular vector (and any vector
in the submodule
generated from it) has zero length and should be removed in order to get a
well-defined Hilbert space.  Not all Verma modules $V_{(c,h)}$ have singular
vectors.  They arise in the Verma modules whose pair $(c,h)$ lies on
at least one of the following curves $(c(t), h_{p,q}(t))$ in the $(c,h)$-plane:
$$\aligned
c(t)&=\frac{15}2 + 3t^{-1}+3t\\
h_{p,q}(t)&={1-p^2\over 8}t^{-1}+{1-pq\over 4}+{1-q^2\over 8}t,\\
\endaligned\tag2.4
$$
where $t\in\Bbb C$ and $p,q\in \Bbb N$ with $p=q \mod 2$.  If $(c,h)=(c(t),
h_{p,q}(t))$ for some $t$, $V_{(c,h)}$ will have a singular vector of weight
$h+{pq\over 2}$.  (See Kac \cite1.) Though not conventional in the physics
literature, the parametrization (2.4) is appropriate to
understand that singular
vectors are not restricted to unitary modules (the very special case $t=-{m
\over m+2}$ with $m=2,3,\dots$) and that the formulae for singular vectors
that follow apply to all cases.

Explicit expressions for singular vectors are known only in the special cases
when either $p$ or $q$ is equal to 1.\cite5  If $p=1$, the singular vector
$\vert \psi_{1,q}\rangle\in V_{(c(t), h_{1,q}(t))}$ is
of weight $h_{1,q}(t)+q/2$
and takes the following form:
$$\aligned
\vert\psi_{1,q}\rangle = \sum\Sb{\text{partitions}}\{k_1,k_2,\dots,k_N\}\\
{\text{of}}\,\,q\,\, {\text{into odd integers}}
\endSb\ \sum\Sb{\text{permutations}}\ \sigma\\
{\text{of}}\,\,N{\text{elements}}\endSb
(\tfrac{t}{2})^{{\frac{q-N}{2}}}\,&
c\,[ k_{\sigma(1)},k_{\sigma(2)},\dots,k_{\sigma(n)}] \\ &\times \ \vert
k_{\sigma(1)}k_{\sigma(2)}\dots k_{\sigma(N)}\rangle\endaligned\tag2.5
$$
where
$$
\align
\vert k_1,k_2,\dots,k_N\rangle
         &= G_{-\tfrac{k_1}{2}}G_{-\tfrac{k_2}{2}} \dots
G_{-\tfrac{k_N}{2}}\vert h\rangle,\\
c\,[k_1,k_2,\dots,k_N]
         &=\prod_{i=1}^N \binom{k_i - 1}{(k_i -1)/2}
\prod_{j=1}^{(N -1)/2}\left(\frac{2}{\sigma_{2j}}
\cdot \frac{2}{\rho_{2j}} \right),\\\endalign
$$
$$
\sigma_j=\sum_{\ell=1}^j k_\ell\qquad{\text{and}}\qquad
\rho_j   =\sum_{\ell=j}^N k_\ell.
$$
The singular vectors are normalized so that
$c[\overbrace{1,1,\dots,1}^{\text{$q$ times}}]=
\left[\left(\frac{q-1}{2}\right)!\right]^{-2}$.
This requirement insures that the above expression never
vanishes.  Let us stress that this form depends only on the $G_{-r}$ (some
$G_{-r}$'s being repeated) and consequently that this form is not in the usual
basis (2.3).

In the Virasoro case, singular vectors had their importance revealed in the
original paper by Belavin, Polyakov and Zamolodchikov \cite6.
The secondary field $\psi(z)=\sum c[k_1, \dots, k_M]
\Cal L_{-k_1}(z)\dots \Cal L_{-k_M}(z)\phi_h(z)$ which
creates out of the vacuum
the singular vector $\vert \psi\rangle=\sum c[k_1, \dots, k_M]
\allowbreak L_{-k_1}\dots L_{-k_M}\vert
h\rangle$\footnote{
As usual, we define $\vert h\rangle\equiv\phi_h(0)\vert 0\rangle$ and the
operators $\Cal L_k(z)$ are related to the $L_k$ by $\Cal L_k(0)=L_k$.
The primary field $\phi_h(z)$, the secondary fields and
their linear combinations constitute the conformal family $[\phi_{(c,h)}(z)]$
which is
in itself
a highest weight representation of the Virasoro algebra.}
belonging to the Verma modula $V_{(c,h)}$
should have zero correlation functions
with any field in the theory.  Even though 2- and 3-point
 correlation functions
are determined by conformal invariance up to a constant, $N$-point ones,
$N\ge 4$, are not and the vanishing of the correlation
functions of the singular
field $\phi(z)$ gives differential equations for them.  These differential
equations are obtained through the isomorphism between
the subalgebra of the
Virasoro algebra spanned by the $L_{-n}$'s with $n\ge 1$
and the subalgebra generated by the following differential
operators $\Cal L_{-n}$.
If one is computing
the $(N+1)$-point correlation function $\langle \psi(z_0)\phi_1(z_1)
\dots\phi_N(z_N)\rangle$ where the $\phi_i(z_i)$ are primary, the differential
equation reads
$$\sum  c[k_1, \dots, k_N] \widehat{\Cal L}_{-k_1}\dots
\widehat{\Cal L}_{-k_N}\langle
\phi(z_0)\phi_1(z_1)\dots\phi_N(z_N)\rangle=0$$
where the $\widehat{\Cal L}_{-n}$ are the differential operators
$$\widehat{\Cal L}_{-n}=\sum_{i=1}^N\left\{ {(k-1)h_i\over z^n_{i0}}-
{1\over z^{n-1}_{i0}}{\partial\over\partial z_{i0}}\right\}$$
with $h_i$ the conformal weight of the field
$\phi_i(z_i)$ and $z_{i0}=z_i-z_0$.
It is easy to check that $[\widehat{\Cal L}_m,
\widehat{\Cal L}_n]= (m-n)\widehat{\Cal L}_{m+n}$.
The analogues of these differential operators for both the $\Cal L_{-n}$'s and
the $\Cal G_{-r}$'s where already given in one of the
earliest works on superconformal
field theories \cite7:
$$\align
\widehat{\Cal L}_{-n}&=\sum_{i=1}^N \left\{ {n-1\over z^n_i}
(h_i +\demi \theta_i\partial_{\theta_i})-{1\over z^{n-1}_i}
\partial_{z_i}\right\}\\
\widehat{\Cal G}_{-r}&=\sum_{i=1}^N \left\{ {1\over z^{r-{1\over 2}}_i}
(\theta_i\partial_{z_i}-\partial_{\theta_i})+
{(1-2r)h_i\over z^{r+{1\over 2}}_i}\theta_i\right\}.\\
\endalign
$$
In these expressions, both $z_0$ and $\theta_0$ have been set to zero.
 We shall
need however the expression for general $z_0$ and $\theta_0$.  Since
correlation functions are invariant under supertranslations both
the correlation
functions and the differential operators $\widehat{\Cal L}_{-n}$'s
and $\widehat{\Cal G}_{-r}$'s
should be expressible in terms of supertranslational invariants.  For the
$(N+1)$-tuplet $(Z_0, Z_1, \dots , Z_N)$, a possible set is known to be:
$$\left. {\aligned \zij i0&=z_i -z_0-\theta_i\theta_0\\
                \theta_{i0}&=\theta_i-\theta_0 \endaligned} \qquad \right\}
\qquad 1\ge i\ge N.\tag2.6$$
It is then easy to write down the general differential operators
 $\widehat{\Cal L}_{-n}$
and $\widehat{\Cal G}_{-r}$:
$$\aligned
\widehat{\Cal L}_{-n}(Z_0; Z_1, \dots, Z_N)=\sum_{i=1}^N\left\{
{(n-1)\over \zij i0^n}(h_i+\demi\tij i0\partial_{\tij i0})
-{1\over\zij i0^{n-1}}\partial_{\zij i0}\right\}, \qquad n\ge 1\\
\widehat{\Cal G}_{-r}(Z_0; Z_1, \dots, Z_N)=\sum_{i=1}^N\left\{
{(1-2r)h_i\over \zij i0^{r+{1\over 2}}}\theta_{i0}
-{1\over \zij i0^{r-{1\over2}}}\left(
\partial_{\theta_{i0}}- \theta_{i0}\partial_{
\zij i0}\right)\right\},\qquad r\ge {1\over 2}.\\
\endaligned\tag2.7
$$

Finaly,it is worth noticing that throughout the rest of the paper we will
use the same notation ($L$ and $G$) for two essentialy different but related
operators;
those acting on a field of the
conformal family $[\phi_{(c,h)}(Z)]$, i.e.~the ${\Cal L}_{-n}(Z)$'s and the
$\Cal G_{-r}(Z)$'s, and those acting on the vectors of its associated
Verma module $V_{(c,h)}$, i.e.~the $L_{-n}$'s and the $G_{-r}$'s. The context
will unambigously indicates which one we should use.

\bigskip\bigskip
\goodbreak
\subhead III. FUSION AND SINGULAR VECTORS\endsubhead
\nobreak\medskip\nobreak
One of the key arguments
of the BdFIZ algorithm is that, since singular fields decouple completely
from the theory, fusion between primary fields whose descendants
 include singular
fields should carry this information to the expanded product.
Among other things
the fusion of a singular field with a primary one should be identically zero.
This idea is rather straightforward though it does not indicate how to extract
new singular vectors out of known ones.  In fact it does so in a very subtle
way and it is the goal of this section to present the BdFIZ algorithm and to
extend it to supersymmetric conformal field theory.

The fusion of two primary superfields $\phi_0(Z_0)$ and $\phi_1(Z_1)$ of
conformal weights $h_0$ and $h_1$ is usually written as:
$$\phi_0(Z_0)\times\phi_1(Z_1)=\sum_h\zij 01^{h-h_0-h_1}
g(h; h_0, h_1) \sum \Sb n\in\Bbb Z\\ \alpha=0,1\endSb \zij
01^n\theta_{01}^\alpha
\phi_h^{(n+{\alpha\over 2})}(Z_1).\tag3.1
$$
The first sum runs over all conformal families $h$ that is contained in the
fusion of $\phi_0$ and $\phi_1$.  Since we will be dealing with only
one of these families at a time, we will select one of these and note by
$\Cal F_h(\phi_0(Z_0)\times\phi_1(Z_1))$ the restriction of the rhs to the
family $h$.  The constant $g(h; h_0, h_1)$ is an overall factor that depends,
besides the conformal weight of the fields involved, on the sectors it is
interpolating from and to.  While these constants $g$ are important in the
normalization of the quantum group action coefficients, they
are irrelevant for the present purpose.  They will be absorbed in the fields
$\phi_h^{(n+{\alpha\over 2})}(Z_1)$ which are descendants of weight
$h+n+{\alpha\over2}$ of the primary field $\phi_h(Z_1)=\phi_h^{(0)}(Z_1)$.
Let us recall, once
for all, that the equality (3.1) is to be understood inside a correlation
function.

As said earlier, the decoupling of a singular field $\psi$ is equivalent to
the vanishing of their fusion with primary fields.  In the notation just
introduced this is simply
$$\Cal F_h(\psi(Z_0)\times\phi_1(Z_1))=0\tag3.2$$
for all $h$.  (This condition seems extremely stringent as it sets all
$\phi_h^{(n+{\alpha\over 2})}(Z)$ in the expansion to zero.) Recall that,
 with
$\psi=\psi_{1,q}$ for example, this condition has the form
$$\align
0&=\Cal F_h(\psi_{1,q}(Z_0)\times\phi_1(Z_1))\\
&=\sum \left({t\over2}\right)^{(q-N)/2}c[k_{\sigma(1)}\dots k_{\sigma(N)}]\\
&\qquad\qquad\times\Cal F_h(G_{-k_1/2}(Z_0) \dots G_{-k_N/2}(Z_0)\phi_0(Z_0)
\times\phi_1(Z_1))
\endalign
$$
where we have used (2.5) and abbreviated the sums over partitions and
permutations by a single sum sign. Can we express
$\Cal F_h\big(\big(G_{-k_1/2} \dots G_{-k_N/2}\phi_0\big)(Z_0)
\times\phi_1(Z_1)\big)$ in terms of $\Cal F_h(\phi_0(Z_0)\times \phi_1(Z_1))$?
This is surely possible since the fusion takes place inside a correlation
function and we know how to take the generators $G_{-k/2}$ out of it by
using the differential operators $\widehat{\Cal G}_{-k/2}$ introduced in (2.7).
Hence we can write
$$\align \Cal F_h\big(\big(G_{-k_1/2} &\dots G_{-k_N/2}\phi_0\big)(Z_0)
\times\phi_1(Z_1)\big) \\
&= \Cal F_h(G_{-k_1/2}\times 1)\dots  \Cal F_h(G_{-k_N/2}\times 1)
\Cal F_h(\phi_0(Z_0)\times \phi_1(Z_1))\endalign
$$
for some operators $\Cal F_h(G_{-k/2}\times 1)$ that we now construct.
Since the
correlation function $\langle \phi_{0}(Z_0)\phi_1(Z_1) \phi_2(Z_2)\dots
\phi_N(Z_N)\rangle$ after fusion will look like
$$\sum_{n,\alpha} \zij 01^{h-h_0-h_1+n}\theta_{01}^\alpha
\langle\phi_h^{(n+{\alpha\over 2})}(Z_1) \phi_2(Z_2)\dots\phi_N(Z_n)\rangle,$$
we want to write $\Cal F_h(G_{-k/2}\times 1)$ as a sum of differential terms
acting on $\zij 01$ and $\theta_{01}$ and of operators
 $\widehat{\Cal G}_{-r}(Z_1;
Z_2, \dots, Z_N)$ and $\widehat{\Cal L}_{-n}(Z_1;
Z_2, \dots, Z_N)$ that could be brought back inside the correlation function
in the form of $G_{-r}$ and $L_{-n}$ acting on $\phi_h^{(n+{\alpha\over2})}$.
Notice that the expression for the
operator $\widehat{{\Cal G}}_{-k/2}(Z_0; Z_1, \dots, Z_N)$
is for a $G_{-k/2}(Z_0)$ acting
on the field $\phi_{0}(Z_0)$ as the operators
 $\widehat{\Cal G}_{-k/2}(Z_1; Z_2,
\dots, Z_N)$ will be for a $G_{-k/2}(Z_1)$ acting on the field $\phi_h^{(n+
{\alpha\over2})}(Z_1)$.  The independent variables used to describe the
functions $\langle \phi_0(Z_0) \dots \phi_N(Z_N)\rangle$ were
$\zij i0$ and $\theta_{i0}$, $1\le i\le N$, and the ones used for
$\langle \phi_h^{(n+{\alpha\over2})}(Z_1) \dots \phi_N(Z_N)\rangle$
should be $\zhat=\zij 01$,
$\theta=\theta_{01}$ and $\zij i1$ and $\theta_{i1}$, $2\le i\le N$.  (In the
latter set of variables, we have labelled the variables
$\zij 01 $ and $\tij 01$
as $\zhat$ and $\th$ respectively, to avoid any confusion.)
{}From here on, it is a simple exercise in many variable calculus.
The new variables are related to the old ones by
$$\aligned
\theta&=-\tij 10,\\
\zhat&=-\zij 10,\\
\tij i1&=\tij i0 - \tij 10,\\
\zij i1&=\zij i0 - \zij 10 -\tij i0 \tij 10, \quad i\ge 2
\endaligned\tag3.3
$$
and the partial derivatives (old ones in terms of new ones):
$$\aligned
\partial_{\tij 10}&=-\partial_{\th}-\sum_{i\ge 2} \partial_{\tij i1}+
\sum_{i\ge 2}(\tij i1-\th)\partial_{\zij i1},\\
\partial_{\zij 10}&=-\partial_{\zhat}-\sum_{i\ge 2}\partial_{\zij i1},\\
\partial_{\tij i0}&=\partial_{\tij i1}+\th \partial_{\zij i1},\\
\partial_{\zij i0}&=\partial_{\zij i1}, \qquad\quad i\ge 2.
\endaligned\tag3.4
$$
After this preamble, the rest of the calculation is straightforward:
$$\align
\Cal F(G_{-r}&(Z_0)\times1)\\&= {1\over (-\zhat)^{r+\demi}} \Big\{
(2r-1)h_1\th - \zhat (\partial_\th + \th \partial_\zhat)
+\zhat(G_{-\demi}(Z_1)+2\th L_{-1}(Z_1))\Big\} \\
&\qquad\qquad+\sum_{m=0}^\infty \binom{r+m-{3\over2}}m \zhat^m(G_{-m-r}(Z_1)+
2\th L_{-m-r-\demi}(Z_1)).\tag3.5a
\endalign
$$
The operator $\Cal F(L_{-n}(Z_0)\times 1)$ is obtained in the same way:
$$\align
\Cal F(L_{-n}&(Z_0)\times1)\\&= {1\over (-\zhat)^{n}}\Big\{
(n-1)h_1 + \demi(n-1)\th(\partial_\th-G_\demi(Z_1))
+\zhat(L_{-1}(Z_1)-
\partial_\zhat)\Big\}\\
&+\sum_{m=0}^\infty \binom{n+m-2}m \zhat^m(L_{-m-n}(Z_1)-
\demi(n+m-1)\th G_{-m-n-\demi}(Z_1)).\tag3.5b\endalign
$$
The $r=\demi$ and $n=1$ cases are the limits of these expressions; they are:
$$\Cal F(G_{-\demi}(Z_0)\times 1)=\th \partial_\zhat + \partial_\th
\qquad{\text{and}}\qquad \Cal F(L_{-1}(Z_0)\times 1)=\partial_\zhat.\tag3.5c$$
One could check that the operators $\Cal F(G_{-r}(Z_0)\times1)$ and
$\Cal F(L_{-n}(Z_0)\times1)$ satisfy the commutation rules of the
subalgebra $\{G_{-r},\ L_{-n}, n,r\geq0\}$ of the Neveu-Schwarz algebra.
Since these expressions do not depend on $h$, we have dropped the index on
$\Cal F_h$.

We are now ready to examine eq.~(3.2), i.e.~what is the content of the fusion
of singular fields with primary ones.  To be more specific, we concentrate
our efforts on the singular vectors $\vert\psi_{1,q}\rangle$.
Let $M_{1,q}$ be the generators
that act on the highest weight vector to create the singular vector
$\vert\psi_{1,q}\rangle$:
$$M_{1,q}=\sum \left({t\over2}\right)^{(q-N)/2} c[k_{\sigma(1)}, \dots,
k_{\sigma(N)}] G_{-k_{\sigma(1)}/2}\dots G_{-k_{\sigma(N)}/2}.$$
Then the decoupling condition is
$$\Cal F(M_{1,q}\times 1) \Cal F_h(\phi_0(Z_0)\times \phi_1(Z_1))=0.
\tag3.6$$
It is easy to compute the action of one term $\Cal F(G_{-r}\times 1)$ on the
fusion of $\phi_0$ and $\phi_1$ to understand the form the above
equation will take. Defining
$$H=h-h_0-h_1,$$
we get
$$\align
\Cal F(&G_{-r}(Z_0)\times1)\big(\zhat^H\sum_{n\ge 0}\zhat^n(\phi_h^{(n)}+
     \th\phi_h^{(n+\demi)})\big)\\
&=\zhat^{H-r-\demi}\sum_{n\ge0}\zhat^n\bigg\{\Big[
     (-1)^{r+\demi}(-\phi_h^{(n-\demi)}+G_{-\demi}\phi_h^{(n-1)})\\
&\qquad\qquad\qquad\qquad\qquad+\sum_{m=0}^{n-r-\demi}\binom{r+m-{3\over2}}m
     G_{-m-r}\phi_h^{(n-m-r-\demi)}
     \Big]\\
&\quad+\th\Big[ (-1)^{r+\demi}((2r-1)h_1-(H+n))\phi_h^{(n)}-
     G_{-\demi}\phi_h^{(n-\demi)}+2L_{-1}\phi_h^{(n-1)}\\
&\quad+\sum_{m=0}^{n-r-\demi}\binom{r+m-{3\over2}}m
     \big(-G_{-m-r}\phi_h^{(n-r-m)}
     +2L_{-m-r-\demi}\phi_h^{(n-r-m-\demi)}\big)\Big]\bigg\}\\
&=\zhat^{H-r-\demi}\sum_{n\ge 0}
\zhat^n(\lambda_h^{(n)}+\th \lambda_h^{(n+\demi)}
     )
\endalign
$$
where
$$\align
\lambda_h^{(n)}&=
\bar N_{n-\demi}
\phi_h^{(n-\demi)}+\sum_{s=1}^{2(n-\demi)}A^1_{s\over2}
    \phi_h^{(n-{s\over2}-\demi)}\\
\lambda_h^{(n+\demi)}&=
\bar N_n
\phi_h^{(n)}+\sum_{s=1}^{2n}A^0_{s\over2}
    \phi_h^{(n-{s\over2})}.
\endalign
$$
The coefficients $A^0_{s\over2}$ and $A^1_{s\over2}$ are functions of
level $s\over2$ of the operators $G_{-r}$ and $L_{-m}$.
The coefficient $\bar N_{n-\demi}$ and $\bar N_n$ of the highest level field,
which will turn out to be crucial in the next lines,
are generated from the part of $\Cal F(G_{-r}(Z_0)\times1)$
that does not depend on the operators $G_{-r}(Z_1)$ and $L_{-n}(Z_1)$:
$$\align
\bar N_{n-\frac\alpha2}&=\zhat^{-(H+n-r-\demi)}{1\over (-\zhat)^{r+\demi}}
\left\{(2r-1)h_1\th - \zhat(\partial_\th +\th\partial_\zhat)\right\}
\zhat^{H+n}\th^\alpha\\
&\equiv\zhat^{-(H+n-r-\demi)}{\hat g}_{-r}(\zhat,\theta)\zhat^{H+n}\th^\alpha,
 \qquad\qquad\qquad\quad \alpha=0,1.\tag3.7
\endalign$$
To get the decoupling equation (3.6), we have to apply several other
 generators
$\Cal F(G_{-r}\times1)$ and gather them into $\Cal F(M_{1,q}\times1)$.
The explicit equations might be long to write down but they will have the form
$$N_{n,\alpha}
(H,h_1)
\phi_h^{(n+{\alpha\over2})}+\sum_{s=1}^{2n+\alpha}
B_{s\over2}\phi_h^{(n+{\alpha\over2}-{s\over2})}=0\tag3.8$$
for $n\ge 0$ and $\alpha=0,1$.  Again $B_{s\over2}$ is a function of the
$G_{-r}$ and $L_{-m}$ with level ${s\over2}$.
Comparing with (3.7), one can see
that the coefficients $N_{n,\alpha}$ of the highest level field
may be put in a compact form as:
$$\th^{1-\alpha}N_{n,\alpha}(H,h_1)=\zhat^{-(H+n+\alpha-q-\demi)}
M_{1,q}\Big(G_{-r}\rightarrow {\hat g}_{-r}(\zhat,\theta)\Big)
\zhat^{H+n}\th^\alpha \tag3.9$$
where the arrow means that all generators $G_{-r}$ in $M_{1,q}$ have to be
replaced by the expression of ${\hat g}_{-r}(\zhat,\theta)$
defined
implicitly
in (3.7).
The next section will be devoted to the
proof of the following identity for $N_{n,\alpha}(H,h_1)$:
$$N_{n,\alpha}(H,h_1)=\prod_{-j+{\alpha\over2}\le M\le j-{\alpha\over2}}
(h_{p',q'}+n+{\alpha\over2}-h_{P,Q+4M})\tag3.10$$
where:
$$h=h_{p',q'},\qquad h_0=h_{1,q}, \qquad h_1=h_{P,Q} \quad {\text{and}}
\quad j=(q-1)/4.$$
The notation (2.4) is being used.

To sum up, the decoupling equations of the singular vector $\phi_{1,q}$
in the conformal family $[\phi_{h_{1,q}}]$ are recursive equations for the
fields $\phi_h^{(n)}$ and $\phi_h^{(n+\demi)}$ of the fusion
$\Cal F_h(\phi_0(Z_0)\times\phi_1(Z_1))$, with $h=h_{1,q}$. This
is of course if the numerical constant $N_{n,\alpha}$ is {\sl not} zero.
In fact, the very first $N_{n=0, \alpha=0}$ must be zero
since otherwise by (3.8) $\phi_h^{(0)}=0$ and
the fusion $\Cal F_h(\phi_0(Z_0)\times\phi_1(Z_1))=0$:
$$\prod (h_{p',q'}-h_{P,Q+4M})=0.\tag3.11$$
In other words, if $N_{n=0, \alpha=0}\ne 0$, the equation (3.8)
$$N_{0,0}\phi_h^{(0)}=0$$
simply says that the conformal family $[\phi_h]$ does not arise in the
fusion of $\phi_0$ and $\phi_1$.  Hence the condition $N_{n=0,
\alpha=0}=0$ is a fusion criterion. Using (3.10), one concludes that the
family $[\phi_{h_{p',q'}}]$ appears in the fusion of $[\phi_{h_{1,q}}]$ and
$[\phi_{h_{P,Q}}]$ if and only if there exists an integer $M\in [-j,j]$ such
that $h_{p',q'}=h_{P,Q+4M}$. A sufficient condition is obviously $p'=P$ and
$q'=Q+4M$ for some integer $M\in[-j,j]$.  (For specific values of $t$, the
condition $N_{0,0}=0$ does not exclude more general fusion rules than the
ones predicted in \cite6.) This is the first consequence of the decoupling
equations.

The second consequence is precisely our goal: the construction of new singular
vectors.  What happens if, besides $N_{0,0}$, another $N_{n,\alpha}$ vanishes?
The recursive process breaks down then and the decoupling equations (3.8)
for this particular $n$ and $\alpha$ is:
$$\sum_{s\ge1}^{2n+\alpha} B_{s\over2}\phi_h^{(n+{\alpha\over2}
-{s\over2})}=0.$$
Since the previous fields $\phi_h^{(m+{\beta\over2})}$, $m+{\beta\over2}
<n+{\alpha\over2}$, have been determined recursively in terms of primary
field $\phi_h^{(0)}=\phi_h$, this is a linear combination of various
homogeneous products of generators $G_{-r}$ and $L_{-m}$ (of level $n+{\alpha
\over2}$) acting on the primary field $\phi_h$.  This expression must then
be a new singular vector of level $n+{\alpha\over2}$
unless the lhs is identically zero.
Although this is not ruled out, in all examples calculated the result has
only a finite number of zeros and singularities in the variable $t$.
Since the singular vector is always defined up to an overall factor that can
be chosen to depend on $t$, this zeros and singularities can be eliminated
and the resuling vector is regular and nonzero for all $t$.

If $h=h_{p',q'}$, the singular vector is expected at the level $p'q'/2$. The
condition $N_{n,\alpha}=0$ is then
$$\prod (h_{p',q'}+p'q'/2-h_{P,Q+4M})=
\prod(h_{p',-q'}-h_{P,Q+4M})=0.\tag3.12$$
The index $M$ runs over the set $\{-j, -j+1, \dots j-1,j\}$ if
the product $p'q'$ is even and
over $\{-j+\demi, -j+{3\over2}, \dots, j-{3\over2}, j-\demi\}$ if it is odd.
If one sticks to generic values of $t$, the conditions (3.11) and (3.12)
can be summed up in:
$$\aligned
q'\le q-1, \ \ P=p', \ \ Q\in\{q'-q+1, q'-q+5, \dots,& -q'+q-5,
-q'+q-1\}\\
& {\text{for $p'$ and $q'$ even,}}\\
q'\le q-2, \ \ P=p', \ \ Q\in\{q'-q+3, q'-q+7, \dots,& -q'+q-7,
-q'+q-3\}\\
& {\text{for $p'$ and $q'$ odd.}}\\
\endaligned\tag3.13
$$
Hence, this algorithm allows for the construction of the singular
vector at level
$p'q'/2$ of the Verma module $V_{(c(t), h_{p',q'}(t))}$,
for all $p', q'\in \Bbb
N$, $p'=q'\mod2$, of the \NS algebra.

To close this section, we outline the construction of the singular vector
$\vert\psi_{2,2}\rangle$ of level 2 in the Verma module
$V_{(c(t),h_{2,2}(t))}$.  Here $h=
h_{p',q'}=h_{2,2}$. The smallest value of $q$ allowed by (3.13) is
$q=3$ and the $Q\in\{0\}$.  Hence, if $q$ is chosen to be 3, the only possible
value of $h_1$ (for generic $t$) is $h_{2,0}$. The conformal weights are then:
$$\align
h&=h_{2,2}=-{3\over8}t^{-1}-{3\over4}-{3\over8}t\\
h_0&=h_{1,3}=-{1\over2}-t\\
h_1&=h_{2,0}=-{3\over8}t^{-1}+{1\over4}+{1\over8}t\\
\intertext{with}
H&=h-h_0-h_1=-\demi+\demi t.
\endalign$$
The singular vector $\vert\psi_{1,3}\rangle\in V_{(c(t), h_{1,3}(t))}$ is
known
by (2.5):
$$\vert\psi_{1,3}\rangle=(tG_{-{3\over2}}+G^3_{-\demi})\vert h_{1,3}\rangle.$$
The algebraic generators $G_{-{3\over2}}$ and $G_{-\demi}$ correspond to the
following generators related to the fusion process (eq.~(3.5)):
$$\align
\Cal F(G_{-\demi}\times 1)&=\partial_\th +\th\partial_\zhat\\
\Cal F(G_{-{3\over2}}\times1)&={1\over\zhat^2}
\Big\{ 2h_1\th-\zhat(\partial_\th
+\th\partial_\zhat) + \sum_{k\ge 1}\zhat^k(G_{-k+\demi}+2\th L_{-k})\Big\}.
\endalign$$
Computing $(t\Cal F(G_{-{3\over2}}\times1)+\Cal F(G^3_{-\demi}\times1))
\Cal F_h(\phi_0(Z_0)\times\phi_1(Z_1))$, one gets the following
recursion equations (eq.~(3.8)):
$$\aligned
({t\over2}+\demi-n)\phi_h^{(n+\demi)} &= \sum_{k=1}^{n+1} t G_{-k+\demi}
\phi_h^{(n-k+1)}\\
n(n-2)\phi_h^{(n)} &= t\sum_{k=1}^n (G_{-k+\demi}\phi_h^{(n-k+\demi)}-
2L_{-k}\phi_h^{(n-k)}).\endaligned\tag3.14
$$
Notice that the coeficient $N_{n,\alpha=0}=n(n-2)$ plainly
 exhibits the required
properties $N_{n=0,\alpha=0}=N_{n=2,\alpha=0}=0$. The three
intermediate fields
$\phi_h^{(\demi)}, \phi_h^{(1)}$ and $\phi_h^{({3\over2})}$
are obtained easily:
$$\align
\phi_h^{(\demi)} & = {2t\over t+1}G_{-\demi}\chphi0 \\
\chphi1 &={2t\over t+1} G_{-\demi}^2 \chphi0 \\
\chphi{{3\over2}} &= {2t\over t-1}G_{-{3\over2}}\chphi0 + {4t^2\over t^2-1}
G_{-\demi}^3\chphi0 .\\
\endalign
$$
The recursion breaks down for $\chphi2$; we get instead the expression for the
singular field of level 2:
$$0=t G_{-\demi}\chphi{{3\over2}} + t G_{-{3\over2}}\chphi\demi-
2tL_{-2}\chphi0 -2tL_{-1}\chphi1$$
which yields:
$$\vert\psi_{2,2}\rangle= \left( {4t\over t^2-1}G_{-\demi}^4+
{t+1\over t-1}G_{-\demi}G_{-{3\over2}}+
{t-1\over t+1}G_{-{3\over2}}G_{-\demi}\right)\vert h_{2,2}\rangle .\tag3.15$$
(The relation  (2.1) was used to get rid of the $L_{-n}$.) The singularities
at $t=\pm 1$ can be removed by simply multiplying the whole
expression by $t^2-1$. A direct check ($G_{-\demi}\vert\psi_{2,2}\rangle=
G_{-{3\over2}}\vert\psi_{2,2}\rangle=0$) shows that this expression is a
singular vector for all values of $t$.

\bigskip\bigskip
\subhead IV. THE COEFFICIENT $N_{n,\alpha}(\lambda, \mu)$\endsubhead
\nobreak\medskip\nobreak
The purpose of the present section is to prove the expression for
$N_{n,\alpha}
(\lambda, \mu)$ given in eq.~(3.10):
$$N_{n,\alpha}(\lambda, \mu)
=\prod_{-j+{\alpha\over2}\le M\le j-{\alpha\over2}}
(h_{p',q'}+n+{\alpha\over2}-h_{P,Q+4M})\tag4.1$$
where
$$\aligned
\lambda &=-h_{P,Q}\\
\mu&=h_{1,q}+h_{P,Q}-h_{p',q'}-n\\
j&={q-1\over4}.\endaligned\tag4.2
$$
Since the proof is rather technical, we spell out here the various steps:
\roster
\item"1." the problem of calculating $N_{n,\alpha}(\lambda, \mu)$ is recast
in the algebraic problem of calculating the determinant of a $q\times q$
matrix $C(\lambda)$ whose elements depend on $\lambda$, i.e.:
$N_{n,\alpha}(\lambda,\mu)=\det C(\lambda)$,
\item"2." the determinant $\det C(\lambda)$ is seen to be a polynom of degree
at most $j$ (resp.~$j+\demi$) if $j$ is an integer (resp.~an integer$ +\demi$);
\item"3." the constant term $\det C(\lambda=0)$ is evaluated;
\item"4." the determinant is evaluated for any $\lambda$.
\endroster

The coefficient $N_{n,\alpha}(\lambda,\mu)$ has been defined in the previous
section (eq.~(3.9)) by:
$$\aligned
\th^{1-\alpha}N_{n,\alpha}(\lambda,&\mu)=\zhat^{{q\over2}+\demi+\mu-\alpha}\\
&\qquad\times M_{1,q}\left(G_{-r}\rightarrow
 {1\over (-\zhat)^{r+\demi}}[(1-2r)\lambda\th
-\zhat(\partial_\th+\th\partial_\zhat)]\right) \zhat^{-\mu}\th^\alpha
\endaligned\tag4.3
$$
where we have already used the coefficients $\lambda$ and $\mu$ defined above.
Though it is not necessary for the present argument, it is interesting to note
that the replacement of the odd generators $G_{-r}$ in the expression for
the singular vector $M_{1,q}$ is done with the differential operators
extending the Witt algebra to a supersymmetric algebra:
$$\aligned
l_{-n}(\lambda) &= -z^{-n}(z\partial_z -
{\tsize{\frac{n-1}2}}\th\partial_\th+\lambda
(n-1)), \qquad n\in \Bbb Z\\
g_{-r}(\lambda) &= z^{-r-\demi}(z\partial_\th - \th z\partial_z-
\th\lambda(2r-1)), \qquad r\in \Bbb Z+\demi.\endaligned\tag4.4
$$
The differential generators in (4.3) coincide with $g_{-r}$ if
$\zhat\rightarrow -z$.  Hence, the coefficient
$N_{n,\alpha}(\lambda,\mu)$ can be rewritten as
$$\th^{1-\alpha}N_{n,\alpha}(\lambda, \mu)=(-1)^{{q+1\over2}+\alpha}
z^{{q+1\over2}+\mu-\alpha} M_{1,q}(G_{-r}\rightarrow g_{-r}(\lambda))
z^{-\mu}\th^\alpha .\tag4.5
$$
Since the sign $(-1)^{{q+1\over2}+\alpha}$ is not relevant for the algorithm,
we shall omit it in the rest of the section.  Accordingly, the equalities
for $N_{n,\alpha}$ are to be understood up to a sign.

The following formulation \cite8 for the singular vectors $M_{1,q}$ will be
used instead of eq.~(2.5) to transform the problem
of calculating $N_{n,\alpha}$
into an algebraic problem.  Let $\Phi$ and $\Psi$ be two $q$-component
vectors with the following forms:
$$\Phi=\pmatrix \phi_{q-1}\\ \phi_{q-2}\\ \vdots \\ \phi_1\\ \phi_0
       \endpmatrix  \qquad{\text{and}}\qquad
  \Psi=\pmatrix \phi_q \\ 0 \\ \vdots \\ 0 \\ 0 \endpmatrix.\tag4.6
$$
Each of the components is itself a vector in the Verma module
$V_{(c,h_{p,q})}$, with $\phi_0$ being the highest weight vector $\vert
h_{1,q}\rangle$ and $\phi_q=M_{1,q}\phi_0$ the singular vector at level $q/2$.
Then, the singular vector $\phi_q$ can be obtained recursively through
the algebraic equation:
$$\Psi=B \Phi=\left(-J_-+\sum_{k=0}^{(q-1)/2}\binom{2k}k G_{-k-\demi}
(\nu \jp)^{2k}\right) \Phi, \qquad {\text{with}}\quad \nu=\sqrt{-{t\over2}}
\tag4.7
$$
where:
$$\jm=\pmatrix 0&0&\dots &0&0\\
               1&0&\dots &0&0\\
               0&1&\dots &0&0\\
               0&0&\dots &0&0\\
               \vdots &\vdots&\ddots&\vdots&\vdots\\
               0&0&\dots &0&0\\
               0&0&\dots &1&0\endpmatrix,\
  \jp=\pmatrix 0 & q-1 & 0 & 0 & \dots & 0 & 0 \\
               0 & 0  & -2 & 0 & \dots & 0 & 0 \\
               0 & 0  & 0 & q-3& \dots & 0 & 0 \\
               \vdots&\vdots &\vdots&\vdots&\ddots&\vdots&\vdots\\
               0 &0&0&0&\dots& 2 & 0\\
               0 &0&0&0&\dots& 0 & -q+1\\
               0 &0&0&0&\dots& 0 & 0\endpmatrix.
$$
The matrix elements of $\jp$ are $(\jp)_{ij}={1\over4}((q-i)(1-(-1)^i)-
i(1+(-1)^i))\delta_{i+1,j})$.
Together with $\jo={1\over4} \text{diag} (q-1, q-3, \dots, -q+3, -q+1)$ and
$\jpdeux, \jm^2$, these $q\times q$ matrices
close an $osp(1\vert 2)$ superalgebra:
$$[\jo,J_\pm ]=\pm\demi J_\pm, \qquad \{\jp, \jm\}=2\jo.\tag4.8
$$
Using this formulation, the
definition of $N_{n,\alpha}$ becomes:
$$\th^{1-\alpha}z^{\alpha-\mu-{q+1\over2}}\pmatrix N_{n,\alpha}\\
0\\0\\ \vdots\\ 0\endpmatrix=\left(-\jm + \sum_{k=0}^{(q-1)/2}
\binom{2k}k(\nu\jp)^{2k}g_{-k-\demi}(\lambda)\right)\pmatrix \phi_{q-1}\\
\phi_{q-2}\\ \vdots\\ \phi_1 \\ \th^\alpha z^{-\mu}\endpmatrix.\tag4.9
$$
 From the realization of the $g_{-k-\demi}(\lambda)$ and the structure of the
matrix $B$, one can conclude that the components of $\Phi$ will have the
following dependancy on $\th$ and $z$:
$$\align
\phi_0&=z^{-\mu},\qquad\phi_1=c_1z^{-\mu-1}\th,\qquad\phi_2=c_2z^{-\mu-1},\\
\phi_3&=c_3z^{-\mu-2}\th,\qquad\dots,\qquad
\phi_{q-1}=c_{q-1}z^{-\mu-{q-1\over2}}\qquad
\text{if $\alpha=0$}\\
\intertext{and}
\phi_0&= z^{-\mu}\th,\qquad\phi_1=c_1z^{-\mu},
\qquad\phi_2=c_2 z^{-\mu-1}\th,\\
\phi_3&=c_3z^{-\mu-1}, \qquad\dots,\qquad
\phi_{q-1}=c_{q-1}z^{-\mu-{q-1\over2}}\th\qquad
\text{if $\alpha=1$}\\
\endalign
$$
where the $c_i$'s may depend on $\alpha,\mu$ and
$\lambda$ but on neither $\th$
nor $z$. Because the parity of the components $\phi_i$, $0\le i\le q-1$,
alternate between even and odd, it is natural to introduce the two matrices
$$A=\text{\ diag\ }(1,0,1,\dots,0,1)\qquad \text{and}\qquad
A'=1-A=\text{\ diag\ }(0,1,0,\dots,1,0)$$
to decompose the action of the $g_{-k-\demi}$ on $\Phi$ as
$$g_{-k-\demi}(A+A')\Phi=-\th z^{-k-1}(z\partial_z+2\lambda k)A\Phi
+z^{-k}\partial_\th A'\Phi.$$
if $\alpha=0$.  For $\alpha=1$, the roles of $A$ and $A'$ are simply
interchanged.  If $\alpha=0$,
the first term $-\th z^{-k-1}(z\partial_z+2\lambda k)A\Phi
=-\th z^{-k-1}(-\mu-j-\jo+2\lambda k)A\Phi$ where we have used $j=
(q-1)/4$.  In the case $\alpha=1$, the relation holds with again
$A\leftrightarrow A'$. With these observations, one sees that
for each side of the linear system (4.9)
the line $i$
is a monomial of the form constant$\times
z^{-\mu-(q-i)/2-\beta}\ \theta^{\beta}$, where $\beta=(i+\alpha)\mod2$.
Hence, the $z$ and $\th$ dependence can be factored out leaving

$$\pmatrix N_{n,\alpha}\\ 0\\ 0\\ \vdots\\ 0\endpmatrix=
\bar C \pmatrix c_{q-1} \\ c_{q-2} \\ \vdots \\ c_1\\ 1\endpmatrix$$
whose solution is simply
$$N_{n,\alpha}(\lambda,\mu)=\det \bar C,$$
because $\bar C$ is the sum of an upper triangular matrix and of $-J_-$:
$$\bar C=-\jm+\sum_{k=0}^{(q-1)/2} \binom{2k}k (\nu\jp)^{2k}
[(\jo+\mu+j-2\lambda k)A+A']$$
for $\alpha=0$, and with $A$ and $A'$ interchanged for $\alpha=1$.  From now
on, we will concentrate on the case $\alpha=0$ only giving the answer for
$\alpha=1$. Using the nilpotency of $\jp$, one can transform $\bar C$ into
$$\bar C=-\jm+\monstre^{-\demi}[(\mu+j+\jo)A+A']+2\lambda t
\monstre^{-{3\over2}}\jpdeux.$$
Since $\det\,\monstre^{3\over2}=1$, we define a new matrix $C$ as
$$C=\monstre^{3\over 2}\bar C$$
and
$$\align
N_{n,0}
&(\lambda,\mu)=\det\, C\\
&=\det\,\big\{ -\monstre^{3\over2}\jm + \monstre [(\mu+j+\jo)A+A']
+2\lambda t \jpdeux A\big\}\tag4.10\endalign
$$
and $N_{n,1}=N_{n,0}(A\leftrightarrow A')$.
The calculation of $N_{n,\alpha}(\lambda,\mu)$ is now
a purely algebraic problem.

The second steps consists in showing that $N_{n,0}(\lambda,\mu)$ is
a polynomial in $\lambda$ of degree at most $j$ (resp.~$(j+\demi)$) if $j$ is
an integer (resp.~an integer$+\demi$).  To prove this, observe first that
the variable $\lambda$ appears in $C$ only in the matrix elements of the form
$C_{n,n+2}$ with $1\le n\le q-2$ and $n$ odd.
(For the present argument, the lines and columns are numbered by indices
running from 1 to $q$.) Moreover these matrix elements are linear in
$\lambda$. The determinant is a sum of products $\prod_{i=1}^q C_{i,k_i}$
where $(k_1, k_2, \dots, k_q)$ is a permutation of the first $q$ integers.
  The
second crucial observation is that, if a product contains a pair $C_{n,n+2}
C_{n+2,n+4}$, it must vanish. To see this, let us try to match the
remaining values for the lines:
$$\{1,2,\dots, n-1, n+1,n+3,n+4,n+5,\dots, q\}$$
with the remaining ones for the columns:
$$\{1,2,\dots, n-1,n,n+1,n+3,n+5, \dots, q\}.$$
Because $C$ has non-vanishing entries only on and over the non-zero diagonal
of $\jm$, i.e.~$C_{ij}\allowbreak
=0$ if $i\ge j+2$, then we have to find a one-to-one
match between the line numbers $\{n+3,n+4, n+5,\dots,q\}$ with the
column numbers $\{n+3, n+5, \dots, q\}$, which is clearly impossible.  Hence,
at best, the contributing products $\prod_{i=1}^q C_{i,k_i}$
will skip every other $\lambda$. That ends the proof.
The proof proceeds in the same way for the $\alpha=1$ case where one finds
that $N_{n,1}(\lambda,\mu)$ is a polynomial of degree at most $j$
(resp. ($j-\frac12$)) if $j$ is an integer (resp. an integer$+\frac12$).

As third step, we obtain the constant term in the polynomial $N_{n,\alpha}
(\lambda,\mu)$.  To do so, we multiply the matrix $\bar C$ evaluated at
$\lambda=0$ on the left by
$$B=[1+t(\jm\jpdeux+\jpdeux\jm)A'][A'+A\monstre^\demi]$$
and on the right by
$$B'=[A+A'\monstre^\demi].$$
These two factors have determinant one and do not change the determinant of
$\bar C$.  Using the following properties
$$AJ_\pm=J_\pm A', \qquad A'J_\pm=J_\pm A,\tag4.11a$$
$$[\jm, \monstre^{\pm\demi}]=\pm t \jp\monstre^{-1\pm\demi},\tag4.11b
$$
the determinant takes the simple form:
$$\det\,\bar C=\det\,\big\{-\jm+A'+A(\mu+j+\jo)-
t(\jm\jpdeux+\jpdeux\jm)\jm A\big\}
\tag4.12$$
which is lower triangular.  Using
$$A={[\jp,\jm]+2j+1\over 4j+1}, \tag4.13
$$
one finds:
$$N_{n,\alpha=0}(\lambda=0,\mu)=\prod_{-j\le M\le j} [\mu+j+M+
t(j+M)(2j-2M+1)].$$
Since the properties (4.11) are invariant under the interchange
 $A\leftrightarrow
A'$, equation (4.12) holds also for $\alpha=1$,
provided we interchange $A$ and $A'$ in the matrices
$B$ and $B'$.
Only the last step of the
calculation differs for $\alpha=1$ and the results can be gathered in a single
relation:
$$N_{n,\alpha}(\lambda=0,\mu)=\prod_{-j+{\alpha\over2}\le M\le j
-{\alpha\over2}}
[\mu+j+M+t(j+M)(2j-2M+1)], \qquad\alpha=0,1.\tag4.14
$$

In the fourth and last step, we show that
$$\aligned
N_{n,\alpha}(\lambda,\mu)^2=\prod_{-j+{\alpha\over2}\le M\le j-{\alpha\over2}}
&\Big[\big( \mu+(j+M)(1+t(2j-2M+1))\big)\\
&\times\big(\mu+(j-M)(1+t(2j+2M+1))\big)
-8\lambda M^2 t\Big].\endaligned\tag4.15
$$
Let $n_M$ be the factor in the product corresponding
to the value $M$ of the index.
Then one notes that $n_M=n_{-M}$ and that the square root can be taken
easily.  It is also clear that (4.15) (at $\lambda=0$)
agrees with (4.14). Since the relation (4.15) predicts precisely the upper
limit for the number of zeros of the polynomial $N_{n,\alpha}(\lambda,\mu)$
(see the second step above), the proof of the above relation is reduced
to verifying that
$$\lambda_M={1\over 8tM^2}\big(\mu+(j+M)(1+t(2j-2M+1))\big)
\big(\mu+(j-M)(1+t(2j+2M+1))\big)$$
for $-j+{\alpha\over2}\le M\le j-{\alpha\over2}$, $M\ne 0$, are indeed  zeros
of the polynomial.

To do that, we go back to the matrix $C$ (eq.~(4.10)) which we evaluate at
$\lambda=\lambda_M$ and write in terms of a new variable $\bar\mu$:
$$\mu=-2M\bar\mu -(j+M)(1+(2j+2M+1)t)\qquad\text{for $M\ne 0$},$$
as:
$$\aligned
C(\lambda=\lambda_M)
=-&\monstre^{3\over2}\jm+\monstre[(\jo+j(2\bar\mu+1))A+A']\\
&+\bar\mu(\bar\mu+1+t)A\jpdeux-(j+M)(1+2\bar\mu+(2j+2M+1)t)A.
\endaligned
$$
The goal is now to replace $C$ by

a lower triangular matrix whose determinant is
trivial to calculate.  As we shall see, the following matrix $D$ has this
property:
$$\aligned
D=(1-t(2\jo+2j+1)\jp A')(1-\bar\mu\jp A')&e^{-X} C(\lambda_M)e^X\\
&\times (1-A'\bar\mu \jp)(A+A'(1-2t\jpdeux))\endaligned\tag4.16
$$
for
$$X=-\jpdeux(\bar\mu + 2t(\jo+j))$$
and its determinant is obviously identical to the one of $C$.
To actually
compute $D$, one needs the following identities:
$$\aligned
\jo e^X&=e^X(\jo+X)\\
\jp e^X&=e^{X+t\jpdeux}\jp\\
\jm e^X&= e^{X-t\jpdeux}(J_- -\jp(\bar\mu+2t(\jo+j)))\\
e^{-X}e^{X-t\jpdeux}&=\sqrt{1-2t\jpdeux}\\
e^{-X}e^{X+t\jpdeux}&={1\over\sqrt{1-2t\jpdeux}}.\\
\endaligned\tag4.17
$$
By first conjugating $C(\lambda_M)$ by $e^{-X}$, then multiplying by the first
matrices on each side of the result in (4.16) and then by the two last ones,
 we
get with the use of (4.13):
$$\align
D=-\jm &+A'+\big[ (\jo+j)(1+2\bar\mu+t(2\jo+2j+1))\\
& -(M+j)(1+2\bar\mu+t(2M+2j+1))\big] A.\endalign$$
The determinant of this lower triangular matrix is $0$ since one of the
diagonal entries of $\jo$ is precisely $M$.
If $\alpha=1$ then we interchange $A$ and $A'$
 in $C(\lambda=\lambda_M)$ and the
matrix $D(A\leftrightarrow A')$ is obtain through
$$\aligned
(A'+A(1-2t\jpdeux))(1-\bar\mu\jp A)&e^{-X} C(A\leftrightarrow A')e^X\\
&\times (1-A\bar\mu \jp)(1-tA\jp(2\jo+2j)).\endaligned\tag4.18
$$
This ends the proof of (4.15).
It is then simply a matter of replacing, in (4.15), the variables $\lambda$
and $\mu$ by their original values (4.2) to get the desired
expression (4.1).

\bigskip
\bigskip
\subhead V. CONCLUDING REMARKS\endsubhead
\nobreak\medskip\nobreak
The extension of the BdFIZ algorithm to the \NS superalgebra represents some
improvement on the actual computation of singular vectors.  The defining
equations ($G_{-\demi}\vert\psi_{p,q}\rangle=
G_{-{3\over2}}\vert\psi_{p,q}\rangle=0$) represent a linear system of
$P'(pq-1)+P'(pq-3)$ equations into $P'(pq)$ variables. (Here $P'(n)$
represent the number of partitions of the integer $n$ into integers,
the odd ones being all distinct.) The present technique amounts to solve
a $q\times q$ linear system or, equivalently, to solve $q-1$ recursive
equations.  Both algorithms are difficult
to use for, say, $pq\ge 6$ without the use of a computer.

In their work, Bauer {\sl et al} reproduced the subseries of
singular vectors $\vert\psi_{1,q}\rangle$ from the single vector $\vert
\psi_{2,1}\rangle$. We have not been able to reach a similar result
for the \NS subseries.  That might be due to the absence of a singular
vector $\vert\psi_{2,1}\rangle$ for the \NS algebra. There is however such
a singular vector in the representation theory of the Ramond algebra.
It would be instructive to understand the interplay of these two sectors
with respect to the singular vector problem.

\bigskip
\bigskip
\subhead Acknowledgements\endsubhead
\nobreak\medskip\nobreak
L.~B.~gratefully acknowledges a scholarship from the Centre de recherches
math\'e\-ma\-tiques of the Universit\'e de Montr\'eal.
 Y.~S.-A.~wants to thank
for the warm hospitality extended to him at the Institute for Advanced Study
where this work was finished.

\bigskip
\bigskip
\subhead References\endsubhead
\nobreak\medskip\nobreak
%%REFERENC#1#2#3S%%%%%%%%%%%%%%%%%%%%%%%%%%%%%%%%%%%%%%%%%%%%%%%%%%%
\def\PL#1#2#3{{\sl Phys.~Lett.}~{\bf #1} (#3) #2.}
\def\NP#1#2#3{{\sl Nucl.~Phys.}~{\bf #1} (#3) #2.}

\def\LMP#1#2#3{{\sl Lett.~Math.~Phys.}~{\bf #1} (#3) #2.}

\def\IJ#1#2#3{{\sl Int.~Jour.~Mod.~Phys.}~{\bf #1} (#3) #2.}
%%%%%%%%%%%%%%%%%%%%%%%%%%%%%%%%%%%%%%%%%%%%%%%%%%%%%%%%%%%%%%%%%%%%%%%%
\baselineskip 18truept
\roster
\item"[1]" V.G.~Kac, {\sl in} Proc.~of the Int.~Cong.~of Math.~(Helsinky,
1978); Group Theor.~Meth. in Phys., Lecture Notes in Physics,
eds.~W.~Beiglb\"ock, A.~B\"ohm and E.~Takasugi
(Springer Verlag, Berlin, 1979),
Vol.~94, p.~441.

\item"[2]" B.L.~Feigin, D.B.~Fuchs, Verma modules over the Virasoro algebra,
{\sl in} Topology, Lecture Notes in Mathematics, eds.~L.D.~Faddeev and
A.A.~Mal'cev (Springer Verlag, Berlin, 1984), pp.~230-245.

\item"[3]" L.~Benoit, Y.~Saint-Aubin, \PL{B215}{517}{1988}

\item"[4]" M.~Bauer, Ph.~Di Francesco, C.~Itzykson, J.-B.~Zuber,
\NP{B362}{515-562}{1991}

\item"[5]" L.~Benoit, Y.~Saint-Aubin, \IJ{A7}{3023}{1992}

\item"[6]" A.A.~Belavin, A.M.~Polyakov, A.B.~Zamolodchikov,
 \NP{B241}{333}{1984}

\item"[7]" M.A.~Bershadsky, V.G.~Knizhnik, M.G.~Teitelman, \PL{B151}{31}{1985}

\item"[8]" L.~Benoit, Y.~Saint-Aubin, \LMP{23}{117}{1991}

\endroster

\enddocument